\documentclass[
    ,final            
  ,numberedheadings 
  ]
  {aipproc}

\layoutstyle{6x9}

\usepackage[centertags]{amsmath}
\allowdisplaybreaks[4]
\usepackage{amssymb}

\newcommand{\href}[1]{\null}
\renewcommand{\url}[1]{#1}

\begin{document}

\title{Neutrino Flavor States and the Quantum Theory of Neutrino Oscillations}


\author{Carlo Giunti}{
  address={INFN, Sezione di Torino, Via P. Giuria 1, I--10125 Torino, Italy}
}

\begin{abstract}
The standard theory of neutrino oscillations is reviewed,
highlighting the main assumptions:
the definition of the flavor states,
the equal-momentum assumption
and
the time $=$ distance assumption.
It is shown that the standard flavor states are
correct approximations of the states that describe neutrinos in oscillation experiments.
The equal-momentum assumption is shown to be unnecessary for the derivation of
the oscillation probability.
The time $=$ distance assumption derives from the wave-packet character
of the propagating neutrinos.
We present a simple quantum-mechanical wave-packet model
which allows us to describe the coherence and localization of neutrino oscillations.
\begin{center}
XI Mexican Workshop on Particles and Fields
\\
7-12 November 2007, Tuxtla Gutierrez, Chiapas, Mexico
\end{center}
\end{abstract}

\maketitle


\section{Introduction}
\label{Introduction}

The idea of neutrino oscillations was discovered by Bruno Pontecorvo in the late 50s
in analogy with $K^{0}$-$\bar{K}^{0}$ oscillations
\cite{Pontecorvo:1957cp,Pontecorvo:1958qd}.
In essence,
neutrino oscillations are lepton flavor transitions
which depend on the distance and time of neutrino propagation
between a source and a detector.
This is a quantum-mechanical effect due to neutrino mixing,
i.e. the fact that
flavor neutrinos are coherent superpositions of massive neutrinos.
The oscillations are caused by the interference
of the different massive neutrinos,
which have different phase velocities.

Since in the late 1950s only one \emph{active} flavor neutrino was known,
the electron neutrino,
Pontecorvo invented the concept of a \emph{sterile} neutrino $\nu_{s}$ \cite{Pontecorvo:1968fh},
which does not take part in weak interactions.
The muon neutrino was discovered at Brookhaven
in 1962
in the first accelerator neutrino experiment
of Lederman, Schwartz, Steinberger, \textit{et al.}
\cite{Danby:1962nd},
following the independent feasibility estimates of Pontecorvo \cite{Pontecorvo:1959sn} and Schwartz \cite{Schwartz:1960hg}.
Since then,
it became clear that oscillations
between different active neutrino flavors are possible
if neutrinos are massive and mixed\footnote{
In 1962 Maki, Nakagawa, and Sakata
\cite{Maki:1962mu}
considered for the first time a model
with $\nu_{e}$--$\nu_{\mu}$ mixing of different neutrino flavors.
Unfortunately,
this model did not have any impact on neutrino mixing research,
since its existence was unknown to the community until the late 70s \cite{Bilenky:1978nj}.
}.
Indeed,
in 1967
Pontecorvo \cite{Pontecorvo:1968fh} discussed the possibility of a depletion of the solar $\nu_{e}$ flux
due to $\nu_{e} \to \nu_{\mu}$
(or $\nu_{e} \to \nu_{s}$)
transitions before the first measurement in the
Homestake experiment
\cite{Cleveland:1998nv}.
In 1969 Gribov and Pontecorvo \cite{Gribov:1969kq}
discussed solar neutrino oscillations
due to $\nu_{e}$--$\nu_{\mu}$ mixing.

The standard theory of neutrino oscillations
was developed in 1975--76
by Eliezer and Swift
\cite{Eliezer:1976ja},
Fritzsch and Minkowski
\cite{Fritzsch:1976rz},
Bilenky and Pontecorvo
\cite{Bilenky:1976cw,Bilenky:1976yj}.
In this theory, massive neutrinos are treated as plane waves,
having definite energy and momentum.
Such a description, however, is not completely consistent,
because energy--momentum conservation
implies that the creation and detection of massive neutrinos
with definite energies and momenta
is possible only if all the particles involved in the production and detection processes have
definite energies and momenta.
The problem is that in this case
energy--momentum conservation cannot hold simultaneously
for different massive neutrinos
and the production and detection of a superposition of different massive neutrinos
are forbidden.
In order to overcome this problem, it is necessary to treat
neutrinos and the other particles participating in the production
and detection processes as wave packets,
as discussed in section~\ref{h078}.

The plan of this paper is as follows:
in section~\ref{s01}
we review the standard theory of neutrino oscillations,
highlighting the main assumptions,
which are discussed in the following sections;
in section~\ref{h003} we discuss the definition of flavor neutrino states;
in section~\ref{h024} we present a covariant plane-wave theory of neutrino oscillations;
in section~\ref{h078} we discuss the necessity of a wave-packet treatment of neutrino oscillations,
in section~\ref{w01} we present a simple quantum-mechanical wave-packet model of neutrino oscillations,
and finally
in section~\ref{Conclusions} we draw our conclusions.

\section{Standard Theory of Neutrino Oscillations}
\label{s01}

Neutrino oscillations are a consequence of
neutrino mixing:
\begin{equation}
\nu_{\alpha L}(x)
=
\sum_{k} U_{\alpha k} \, \nu_{kL}(x)
\qquad
(\alpha=e,\mu,\tau)
\,,
\label{001}
\end{equation}
where
$\nu_{\alpha L}(x)$
are the left-handed flavor neutrino fields,
$\nu_{kL}(x)$
are the left-handed massive neutrino fields
and
$U$ is the unitary mixing matrix
(see
Refs.~\cite{Bilenky:1978nj,Bilenky:1987ty,hep-ph/9812360,hep-ph/0306239,Giunti-Kim-2007}).
Since
a flavor neutrino $\nu_{\alpha}$
is created by $\nu_{\alpha L}^{\dagger}(x)$
in a charged-current weak interaction process,
in the standard plane-wave theory of neutrino oscillations
\cite{Eliezer:1976ja,Fritzsch:1976rz,Bilenky:1976cw,Bilenky:1976yj,Bilenky:1978nj},
it is assumed that $\nu_{\alpha}$ is described
by the standard flavor state
\begin{equation}
| \nu_{\alpha} \rangle
=
\sum_{k} U_{\alpha k}^* \, | \nu_{k} \rangle
\,,
\label{1331}
\end{equation}
which has the same mixing as the field
$\nu_{\alpha L}^{\dagger}(x)$.

Since the massive neutrino states
$| \nu_{k} \rangle$
have definite mass $m_{k}$ and definite energy $E_{k}$,
they evolve in time as plane waves:
\begin{equation}
i
\,
\frac{\partial}{\partial t}
\,
| \nu_{k}(t) \rangle
=
\mathsf{H}_{0}
\,
| \nu_{k}(t) \rangle
=
E_{k}
\,
| \nu_{k}(t) \rangle
\,,
\qquad
| \nu_{k}(t) \rangle
=
e^{- i E_{k} t}
\,
| \nu_{k} \rangle
\,,
\label{003}
\end{equation}
where $\mathsf{H}_{0}$ is the free Hamiltonian operator,
\begin{equation}
E_{k}^{2} = p_{k}^{2} + m_{k}^{2}
\,,
\label{h029}
\end{equation}
and
$ | \nu_{k}(t=0) \rangle = | \nu_{k} \rangle $
(all the massive neutrinos start with the same arbitrary phase).
The resulting time evolution of the flavor neutrino state in Eq.~(\ref{1331})
is given by
\begin{equation}
| \nu_{\alpha}(t) \rangle
=
\sum_{k} U_{\alpha k}^* \, e^{- i E_{k} t}
\,
| \nu_{k} \rangle
=
\sum_{\beta=e,\mu,\tau}
\left(
\sum_{k} U_{\alpha k}^* \, e^{- i E_{k} t} \, U_{\beta k}
\right)
| \nu_{\beta} \rangle
\,.
\label{004}
\end{equation}
Hence,
if the mixing matrix $U$ is different from unity
(\textit{i.e.} if there is neutrino mixing),
the state $| \nu_{\alpha}(t) \rangle$,
which has pure flavor $\alpha$ at the initial time $t=0$,
evolves in time into a superposition of different flavors.
The quantity in parentheses in Eq.~(\ref{004})
is the amplitude of $\nu_{\alpha}\to\nu_{\beta}$ transitions
at the time $t$ after $\nu_{\alpha}$ production.
The probability of $\nu_{\alpha}\to\nu_{\beta}$ transitions at the time $t=T$
of neutrino detection is given by
\begin{equation}
P_{\alpha\beta}(T)
=
|\langle \nu_{\beta} | \nu_{\alpha}(T) \rangle |^{2}
=
\left|
\sum_{k} U_{\alpha k}^* \, e^{- i E_{k} T} \, U_{\beta k}
\right|^{2}
=
\sum_{k,j} U_{\alpha k}^* \, U_{\beta k} \,  U_{\alpha j} \, U_{\beta j}^*
\, e^{- i ( E_{k} - E_{j} ) T}
\,.
\label{005}
\end{equation}
One can see that
$P_{\alpha\beta}(T)$
depends on the energy differences
$E_{k} - E_{j}$.
In the standard theory of neutrino oscillations
it is assumed that all massive neutrinos have the same momentum $\vec{p}$.
Since detectable neutrinos are ultrarelativistic\footnote{
It is known that neutrino masses
are smaller than about one eV
(see the reviews in Refs.~\cite{hep-ph/0211462,hep-ph/0310238}).
Since only neutrinos with energy larger than about 100 keV
can be detected (see the discussion in Ref.~\cite{hep-ph/0205014}),
in oscillation experiments neutrinos are always ultrarelativistic.
},
we have
\begin{equation}
E_{k}
\simeq
E
+
\frac{m_{k}^{2}}{2E}
\,,
\qquad
E_{k} - E_{j}
=
\frac{\Delta{m}^{2}_{kj}}{2E}
\,,
\label{006}
\end{equation}
where
$\Delta{m}^{2}_{kj} \equiv m_{k}^{2} - m_{j}^{2}$
and
$E \equiv |\vec{p}|$
is the energy of a massless neutrino
(or, in other words,
the neutrino energy in the massless approximation).
In most neutrino oscillation experiments
the time $T$ between production and detection is not measured,
but the source-detector distance $L$ is known.
In this case,
in order to apply the oscillation probability to the data analysis
it is necessary to express $T$ as a function of $L$.
Considering ultrarelativistic neutrinos,
we have $T \simeq L$,
leading to the standard formula for the oscillation probability:
\begin{equation}
P_{\alpha\beta}(L,E)
=
\sum_{k,j}
U_{{\alpha}k}^{*}
\,
U_{{\beta}k}
\,
U_{{\alpha}j}
\,
U_{{\beta}j}^{*}
\,
\exp\left( - i \, \frac{\Delta{m}^{2}_{kj} L}{2E} \right)
\,.
\label{a001}
\end{equation}

Summarizing,
there are three main assumptions in the standard theory of neutrino oscillations:
\renewcommand{\labelenumi}{\theenumi}
\renewcommand{\theenumi}{(A\arabic{enumi})}
\begin{enumerate}
\item
\label{A1}
Neutrinos produced or detected in
charged-current weak interaction processes
are described by the
flavor states in Eq.~(\ref{1331}).
\item
\label{A2}
The massive neutrino states
$|\nu_{k}\rangle$
in Eq.~(\ref{1331})
have the same momentum
(``\emph{equal-momentum assumption}'').
\item
\label{A3}
The propagation time
is equal to the
distance $L$
traveled by the neutrino
between production and detection
(``\emph{time $=$ distance assumption}'').
\end{enumerate}
In the following we will show that the assumptions~\ref{A1} and ~\ref{A3}
correspond to approximations which are appropriate in the analysis of current
neutrino oscillation experiments (section~\ref{h003} and \ref{h078}, respectively).
Instead,
the equal-momentum assumption~\ref{A2} is not physically justified
\cite{Winter:1981kj,Giunti:1991ca,hep-ph/0011074,hep-ph/0104148,hep-ph/0302026},
as one can easily understand from the application of energy-momentum conservation
to the production process\footnote{
A different opinion,
in favor of the equal-momentum assumption,
has been recently expressed in
Ref.~\cite{hep-ph/0604044}.
On the other hand,
other authors \cite{hep-ph/9607201,hep-ph/9802387,hep-ph/0304187}
advocated an equal-energy assumption,
which we consider as unphysical as the equal-momentum assumption.
}.
However,
in section~\ref{h024} we will show that the assumption~\ref{A2}
is actually not necessary for the derivation of the
oscillation probability if both the evolutions in space and in time
of the neutrino state are taken into account.

\section{Flavor Neutrino States}
\label{h003}

The state of a flavor neutrino $\nu_{\alpha}$
is defined as the state
which describes a neutrino produced in a charged-current weak interaction process
together with a charged lepton
$\ell_{\alpha}^{+}$
or from a charged lepton
$\ell_{\alpha}^{-}$
($ \ell_{\alpha}^{\pm} = e^{\pm} , \mu^{\pm} , \tau^{\pm} $ for $\alpha=e,\mu,\tau$, respectively),
or the state
which describes a neutrino detected in a charged-current weak interaction process
with a charged lepton
$\ell_{\alpha}^{-}$ in the final state.
In fact, the neutrino flavor can only be measured
through the identification of the charged lepton
associated with the neutrino
in a charged-current weak interaction process.

Let us first consider a neutrino produced in the generic decay process
\begin{equation}
\text{P}_{\text{I}} \to \text{P}_{\text{F}} + \ell_{\alpha}^{+} + \nu_{\alpha}
\,,
\label{h007}
\end{equation}
where $\text{P}_{\text{I}}$ is the decaying particle
and $\text{P}_{\text{F}}$ represents any number of final particles.
For example: in the pion decay process
$
\pi^{+} \to \mu^{+} + \nu_{\mu}
$
we have $\text{P}_{\text{I}}=\pi^{+}$, $\text{P}_{\text{F}}$ is absent and $\alpha=\mu$;
in a nuclear $\beta^{+}$ decay process
$
\text{N}(A,Z)
\to
\text{N}(A,Z-1)
+
e^{+} + \nu_{e}
$
we have $\text{P}_{\text{I}}=\text{N}(A,Z)$, $\text{P}_{\text{F}}=\text{N}(A,Z-1)$ and $\alpha=e$.
The following method can easily be
modified in the case of a $\nu_{\alpha}$ produced in the generic scattering process
$ \ell_{\alpha}^{-} + \text{P}_{\text{I}} \to \text{P}_{\text{F}} + \nu_{\alpha} $
by replacing the $\ell_{\alpha}^{+}$ in the final state with a $\ell_{\alpha}^{-}$
in the initial state.

The final state resulting from the decay
of the initial particle $\text{P}_{\text{I}}$ is given by
\begin{equation}
| f \rangle
=
\mathsf{S} \, | \text{P}_{\text{I}} \rangle
\,,
\label{h011}
\end{equation}
where $\mathsf{S}$
is the $S$-matrix operator.
Since
the final state $ | f \rangle $ contains all the decay channels of $\text{P}_{\text{I}}$,
it can be written as
\begin{equation}
| f \rangle
=
\sum_{k} \mathcal{A}^{\text{P}}_{\alpha k} \, | \nu_{k} , \ell_{\alpha}^{+} , \text{P}_{\text{F}} \rangle
+
\ldots
\,,
\label{h012}
\end{equation}
where we have singled out the decay channel in Eq.~(\ref{h007})
and we have taken into account that the flavor neutrino $\nu_{\alpha}$
is a coherent superposition of massive neutrinos $\nu_{k}$.
Since
the states of the other decay channels represented by dots in Eq.~(\ref{h012})
are orthogonal to $| \nu_{k} , \ell_{\alpha}^{+} , \text{P}_{\text{F}} \rangle$
and
the states
$| \nu_{k} , \ell_{\alpha}^{+} , \text{P}_{\text{F}} \rangle$
with different $k$s
are orthonormal,
the coefficients $\mathcal{A}^{\text{P}}_{\alpha k}$
are the amplitudes of production of the corresponding state
in the decay channel in Eq.~(\ref{h007}):
\begin{equation}
\mathcal{A}^{\text{P}}_{\alpha k}
=
\langle \nu_{k} , \ell_{\alpha}^{+} , \text{P}_{\text{F}} | f \rangle
=
\langle \nu_{k} , \ell_{\alpha}^{+} , \text{P}_{\text{F}} |
\,
\mathsf{S}
\,
| \text{P}_{\text{I}} \rangle
\,.
\label{h013}
\end{equation}
Projecting the final state in Eq.~(\ref{h012})
over $|\ell_{\alpha}^{+} , \text{P}_{\text{F}} \rangle$
and normalizing,
we obtain the flavor neutrino state
\cite{Giunti:1992cb,hep-ph/0102320,hep-ph/0306239,hep-ph/0402217}
\begin{equation}
| \nu_{\alpha}^{\text{P}} \rangle
=
\left( \sum_{i} |\mathcal{A}^{\text{P}}_{\alpha i}|^{2} \right)^{-1/2}
\sum_{k} \mathcal{A}^{\text{P}}_{\alpha k} \, | \nu_{k} \rangle
\,.
\label{h015}
\end{equation}
Therefore, a flavor neutrino state is a coherent superposition of massive neutrino states
$ | \nu_{k} \rangle $
and
the coefficient
$\mathcal{A}^{\text{P}}_{\alpha k}$
of the $k$th massive neutrino component
is given by the amplitude of production of $\nu_{k}$.
Since, in general,
the amplitudes $\mathcal{A}^{\text{P}}_{\alpha k}$ depend on the production process,
a flavor neutrino state depends on the production process.
In the following,
we will call a flavor neutrino state of the type in Eq.~(\ref{h015}) a
``\emph{production flavor neutrino state}''.

Let us now consider the detection of a flavor neutrino $\nu_{\alpha}$
through the generic charged-current weak interaction process
\begin{equation}
\nu_{\alpha} + \text{D}_{\text{I}} \to \text{D}_{\text{F}} + \ell_{\alpha}^{-}
\,,
\label{h035}
\end{equation}
where $\text{D}_{\text{I}}$ is the target particle
and $\text{D}_{\text{F}}$ represents one or more final particles.
In general,
since the incoming neutrino state in the detection process is a superposition of
massive neutrino states, it may not have a definite flavor.
Therefore,
we must consider the generic process
\begin{equation}
\nu + \text{D}_{\text{I}}
\,,
\label{h035a}
\end{equation}
with a generic incoming neutrino state $ | \nu \rangle $.
In this case,
the final state of the scattering process is given by
\begin{equation}
| f \rangle
=
\mathsf{S} \, | \nu , \text{D}_{\text{I}} \rangle
\,,
\label{h201}
\end{equation}
This final state contains all the possible scattering channels:
\begin{equation}
| f \rangle
=
| \text{D}_{\text{F}} , \ell_{\alpha}^{-} \rangle + \ldots
\,,
\label{h201a}
\end{equation}
where we have singled out the scattering channel in Eq.~(\ref{h035}).
We want to find the component
\begin{equation}
| \nu_{\alpha} , \text{D}_{\text{I}} \rangle
=
\sum_{k} \mathcal{A}^{\text{D}}_{\alpha k} | \nu_{k} , \text{D}_{\text{I}} \rangle
\label{h203}
\end{equation}
of the initial state
$ | \nu , \text{D}_{\text{I}} \rangle $
which corresponds to the flavor $\alpha$,
i.e.\ the component which generates only the scattering channel in Eq.~(\ref{h035}).
This means that
$
| \text{D}_{\text{F}} , \ell_{\alpha}^{-} \rangle
=
\mathsf{S} \, | \nu_{\alpha} , \text{D}_{\text{I}} \rangle
$.
Using the unitarity of the mixing matrix,
we obtain
\begin{equation}
| \nu_{\alpha} , \text{D}_{\text{I}} \rangle
=
\mathsf{S}^{\dagger} \, | \text{D}_{\text{F}} , \ell_{\alpha}^{-} \rangle
\,.
\label{h202}
\end{equation}
From Eqs.~(\ref{h203}) and (\ref{h202}),
the coefficients $\mathcal{A}^{\text{D}}_{\alpha k}$ are the complex conjugate of the
amplitude of detection of $\nu_{k}$ in the detection process in Eq.~(\ref{h035}):
\begin{equation}
\mathcal{A}^{\text{D}}_{\alpha k}
=
\langle \nu_{k} , \text{D}_{\text{I}} | \mathsf{S}^{\dagger} \, | \text{D}_{\text{F}} , \ell_{\alpha}^{-} \rangle
\,.
\label{h204}
\end{equation}
Projecting $ | \nu_{\alpha} , \text{D}_{\text{I}} \rangle $ over $ | \text{D}_{\text{I}} \rangle $ and normalizing,
we finally obtain the flavor neutrino state in the detection process in Eq.~(\ref{h035}):
\begin{equation}
| \nu_{\alpha}^{\text{D}} \rangle
=
\left( \sum_{i} |\mathcal{A}^{\text{D}}_{\alpha i}|^{2} \right)^{-1/2}
\sum_{k} \mathcal{A}^{\text{D}}_{\alpha k} \, | \nu_{k} \rangle
\,.
\label{h205}
\end{equation}
In the following,
we will call a flavor neutrino state of this type a
``\emph{detection flavor neutrino state}''.

Although
the expressions in Eqs.~(\ref{h015}) and (\ref{h205})
for the production and detection flavor neutrino states
have the same structure,
these states have different meanings.
A production flavor neutrino state describes the neutrino created in a
charged-current interaction process, which propagates out of a source.
Hence, it describes the initial state of a propagating neutrino.
A detection flavor neutrino state does not describe a propagating neutrino.
It describes the component of the state of a propagating neutrino which
can generate a charged lepton with appropriate flavor
through a charged-current weak interaction with
an appropriate target particle.
In other words, the scalar product
\begin{equation}
A_{\alpha}
=
\langle \nu_{\alpha}^{\text{D}} | \nu \rangle
\label{h206}
\end{equation}
is the probability amplitude to find a $\nu_{\alpha}$
by observing the scattering channel in Eq.~(\ref{h035}) with the scattering process in Eq.~(\ref{h035a}).

In order to understand the connection of the production and detection flavor neutrino states
with the standard flavor neutrino states in Eq.~(\ref{1331}),
it is useful to express the $S$-matrix operator as
\begin{equation}
\mathsf{S}
=
1
- i \int \text{d}^{4}x \,
\mathsf{H}_{\text{CC}}(x)
\,,
\qquad
\mathsf{H}_{\text{CC}}(x)
=
\frac{ G_{\text{F}} }{ \sqrt{2} }
\,
j_{\rho}^{\dagger}(x) \, j^{\rho}(x)
\,,
\label{h016}
\end{equation}
where $G_{\text{F}}$ is the Fermi constant
(we considered
only the first order perturbative contribution of the
effective low-energy charged-current weak interaction Hamiltonian).
The weak charged current $j^{\rho}(x)$ is given by
\begin{align}
j^{\rho}(x)
=
\null & \null
\sum_{\alpha=e,\mu,\tau}
\overline{\nu_{\alpha}}(x)
\,
\gamma^{\rho}
\left( 1 - \gamma^{5} \right)
\ell_{\alpha}(x)
+
h^{\rho}(x)
\nonumber
\\
=
\null & \null
\sum_{\alpha=e,\mu,\tau}
\sum_{k}
U_{\alpha k}^{*}
\,
\overline{\nu_{k}}(x)
\,
\gamma^{\rho}
\left( 1 - \gamma^{5} \right)
\ell_{\alpha}(x)
+
h^{\rho}(x)
\,,
\label{h017}
\end{align}
where $h^{\rho}(x)$ is the hadronic weak charged current.
The production and detection amplitudes
$\mathcal{A}^{\text{P}}_{\alpha k}$
and
$\mathcal{A}^{\text{D}}_{\alpha k}$
can be written as
\begin{equation}
\mathcal{A}^{\text{P}}_{\alpha k}
=
U_{\alpha k}^{*}
\,
\mathcal{M}^{\text{P}}_{\alpha k}
\,,
\qquad
\mathcal{A}^{\text{D}}_{\alpha k}
=
U_{\alpha k}^{*}
\,
\mathcal{M}^{\text{D}}_{\alpha k}
\,,
\label{h018}
\end{equation}
with the interaction matrix elements
\begin{align}
\null & \null
\mathcal{M}^{\text{P}}_{\alpha k}
=
- i
\,
\frac{G_{\text{F}}}{\sqrt{2}}
\int \text{d}^{4}x \,
\langle \nu_{k} , \ell_{\alpha}^{+} |
\,
\overline{\nu_{k}}(x)
\,
\gamma^{\rho}
\left( 1 - \gamma^{5} \right)
\ell_{\alpha}(x)
\,
| 0 \rangle
\,
J_{\rho}^{\text{P}_{\text{I}} \to \text{P}_{\text{F}}}(x)
\,,
\label{h019}
\\
\null & \null
\mathcal{M}^{\text{D}}_{\alpha k}
=
i
\,
\frac{G_{\text{F}}}{\sqrt{2}}
\int \text{d}^{4}x \,
\langle \nu_{k} |
\,
\overline{\nu_{k}}(x)
\,
\gamma^{\rho}
\left( 1 - \gamma^{5} \right)
\ell_{\alpha}(x)
\,
| \ell_{\alpha}^{-} \rangle
\,
{J_{\rho}^{\text{D}_{\text{I}} \to \text{D}_{\text{F}}}}^{*}(x)
\,.
\label{h036}
\end{align}
Here
$J_{\rho}^{\text{P}_{\text{I}} \to \text{P}_{\text{F}}}(x)$
and
$J_{\rho}^{\text{D}_{\text{I}} \to \text{D}_{\text{F}}}(x)$
are,
respectively,
the matrix elements of the
$\text{P}_{\text{I}} \to \text{P}_{\text{F}}$
and
$\text{D}_{\text{I}} \to \text{D}_{\text{F}}$
transitions.

Using Eq.~(\ref{h018}),
the production and detection flavor neutrino states
can be written as
\begin{align}
\null & \null
| \nu_{\alpha}^{\text{P}} \rangle
=
\sum_{k}
\frac{\mathcal{M}^{\text{P}}_{\alpha k}}{ \sqrt{ \sum_{j} |U_{\alpha j}|^{2} \, |\mathcal{M}^{\text{P}}_{\alpha j}|^{2} } }
\,
U_{\alpha k}^{*}
\,
| \nu_{k} \rangle
\,,
\label{h020a}
\\
\null & \null
| \nu_{\alpha}^{\text{D}} \rangle
=
\sum_{k}
\frac{\mathcal{M}^{\text{D}}_{\alpha k}}{ \sqrt{ \sum_{j} |U_{\alpha j}|^{2} \, |\mathcal{M}^{\text{D}}_{\alpha j}|^{2} } }
\,
U_{\alpha k}^{*}
\,
| \nu_{k} \rangle
\,.
\label{h020b}
\end{align}
These states have a structure
which is similar to the standard flavor states in Eq.~(\ref{1331}),
with the relative contribution of
the massive neutrino $\nu_{k}$ proportional to $U_{\alpha k}^{*}$.
The additional factors
are due to the dependence of
the production and detection processes on the neutrino masses.

In experiments which are not sensitive to the dependence of
$\mathcal{M}^{\text{P}}_{\alpha k}$
and
$\mathcal{M}^{\text{D}}_{\alpha k}$
on the difference of the neutrino masses,
it is possible to approximate
\begin{equation}
\mathcal{M}^{\text{P}}_{\alpha k} \simeq \mathcal{M}^{\text{P}}_{\alpha}
\,,
\qquad
\mathcal{M}^{\text{D}}_{\alpha k} \simeq \mathcal{M}^{\text{D}}_{\alpha}
\,.
\label{h021}
\end{equation}
In this case, since
\begin{equation}
\displaystyle \sum_{k} |U_{\alpha k}|^{2} = 1
\,,
\label{h022}
\end{equation}
we obtain,
up to an irrelevant phase,
the standard flavor neutrino states in Eq.~(\ref{1331}),
which do not depend on the production or detection process.
Hence,
the standard flavor neutrino states
are approximations of the production and detection flavor neutrino states
in experiments which are not sensitive to the dependence of
the neutrino interaction rate
on the difference of the neutrino masses.
All neutrino oscillation experiments have this characteristic:
since the detectable neutrinos are ultrarelativistic,
neutrino oscillation experiments are insensitive to any effect of neutrino masses
in the production and detection processes.
Therefore,
the assumption~\ref{A1} in the standard theory of neutrino oscillations
is correct as an appropriate approximation in the analysis of
neutrino oscillation experiments.

\section{Covariant Plane-Wave Theory of Oscillations}
\label{h024}

In this section we show that the equal-momentum assumption~\ref{A2}
can be avoided by considering not only the time evolution of the neutrino states,
as in the standard theory,
but also their space dependence.

Let us consider a neutrino oscillation experiment
in which $ \nu_{\alpha} \to \nu_{\beta} $
transitions are studied
with a production process of the type in Eq.~(\ref{h007})
and a detection process of the type in Eq.~(\ref{h035}).
In this case,
the produced flavor neutrino $ \nu_{\alpha} $
is described by the production flavor state $ | \nu_{\alpha}^{\text{P}} \rangle $ in Eq.~(\ref{h015}).
If the neutrino production and detection processes
are separated by a space-time interval
$(T,L)$,
the neutrino propagates freely between production and detection,
evolving into the state
\begin{equation}
| \nu (T,L) \rangle
=
e^{ -i \mathsf{E} T + i \mathsf{P} L }
\,
| \nu_{\alpha}^{\text{P}} \rangle
\,,
\label{h027a}
\end{equation}
where
$\mathsf{E}\equiv\mathsf{H}_{0}$ and $\mathsf{P}$
are, respectively, the energy and momentum operators.
This is the incoming neutrino state in the detection process.
The amplitude of the measurable $ \nu_{\alpha} \to \nu_{\beta} $ transitions
is given by the scalar product in Eq.~(\ref{h206}):
\begin{equation}
A_{\alpha\beta}(T,L)
=
\langle \nu_{\beta}^{\text{D}} | \nu (T,L) \rangle
=
\langle \nu_{\beta}^{\text{D}} |
e^{ -i \mathsf{E} T + i \mathsf{P} L }
| \nu_{\alpha}^{\text{P}} \rangle
\,,
\label{h027}
\end{equation}
with the detection flavor state
$ | \nu_{\beta}^{\text{D}} \rangle $
in Eq.~(\ref{h205}).

Neglecting mass effects in the production and detection processes,
we approximate the production and detection flavor states with the standard
ones given in Eq.~(\ref{1331}).
Then, we obtain
\begin{equation}
A_{\alpha\beta}(T,L)
=
\sum_{k}
U_{\alpha k}^{*}
\,
U_{\beta k}
\,
e^{ - i E_{k} T + i p_{k} L }
\,.
\label{h030}
\end{equation}
Notice that the consideration of
the space-time interval
between neutrino production and detection allows one to
take into account both the differences
in energy and momentum of the massive neutrinos
\cite{Winter:1981kj,Giunti:1991ca,hep-ph/0011074,hep-ph/0104148,hep-ph/0302026}.

In oscillation experiments in which
the neutrino propagation time $T$ is not measured,
it is possible to adopt the light-ray $T = L$ approximation (assumption~\ref{A3}),
since neutrinos are ultrarelativistic
(the effects of possible deviations from $T = L$
are shown to be negligible in Refs.~\cite{hep-ph/0608070,Giunti-Kim-2007}).
In this case,
the phase in Eq.~(\ref{h030}) becomes
\begin{equation}
- E_{k} T + p_{k} L
=
- \left( E_{k} - p_{k} \right) L
=
- \frac{ E_{k}^{2} - p_{k}^{2} }{ E_{k} + p_{k} } \, L
=
- \frac{ m_{k}^{2} }{ E_{k} + p_{k} } \, L
\simeq
- \frac{ m_{k}^{2} }{ 2 E } \, L
\,,
\label{h032}
\end{equation}
which leads to the standard oscillation probability in Eq.~(\ref{a001}).

Equation~(\ref{h032})
shows that the phases in the flavor transition amplitude
are independent from the values of the energies and momenta
of different massive neutrinos
\cite{Winter:1981kj,Giunti:1991ca,hep-ph/0011074,hep-ph/0104148,hep-ph/0302026},
because of the relativistic dispersion relation in Eq.~(\ref{h029}).
In particular, Eq.~(\ref{h032}) shows that
the equal-momentum assumption~\ref{A2} in section~\ref{s01},
adopted in the standard derivation of the neutrino oscillation probability,
is not necessary
in an improved derivation which takes into account
both the evolutions in space and in time
of the neutrino state.

We have called this derivation of the flavor transition probability
``covariant plane-wave theory of oscillations''
because it is manifestly Lorentz invariant.
This is important because
flavor, which is the quantum number that distinguishes different types
of quarks and leptons,
is a Lorentz-invariant quantity.
For example,
an electron is seen as an electron by any observer,
never as a muon.
Therefore,
the probability of flavor neutrino oscillations
must be Lorentz invariant
\cite{hep-ph/0011074,physics/0305122}.

\section{Wave-Packet Treatment}
\label{h078}

So far, we have considered massive neutrinos as particles
described by plane waves
with definite energy and momentum.
However,
the $T = L$ assumption~\ref{A3}
requires a wave packet description.
The reason is simple:
since
plane waves cover all space-time in a periodic way
they cannot describe the localized events of neutrino production and detection.
As discussed in introductory books on optics
(see \cite{Born-Wolf-PrinciplesOfOptics-1959,Jenkins-White-FundamentalsOfOptics-1981})
and quantum mechanics
(see \cite{Schiff-QuantumMechanics,Bohm-QuantumTheory}),
real localized particles are described by
superpositions of plane waves known as \emph{wave packets}.

Moreover,
different massive neutrinos can be produced and detected coherently
only if the energies and momenta
in the production and detection processes
have sufficiently large uncertainties
\cite{Kayser:1981ye,hep-ph/9506271}.
The uncertainty of the production process implies that the massive neutrinos
propagating between production and detection have a momentum distribution
\cite{hep-ph/0205014},
i.e.\ they are described by wave packets.

\begin{figure}[t!]
\includegraphics*[bb=84 384 511 772, width=0.45\textwidth]{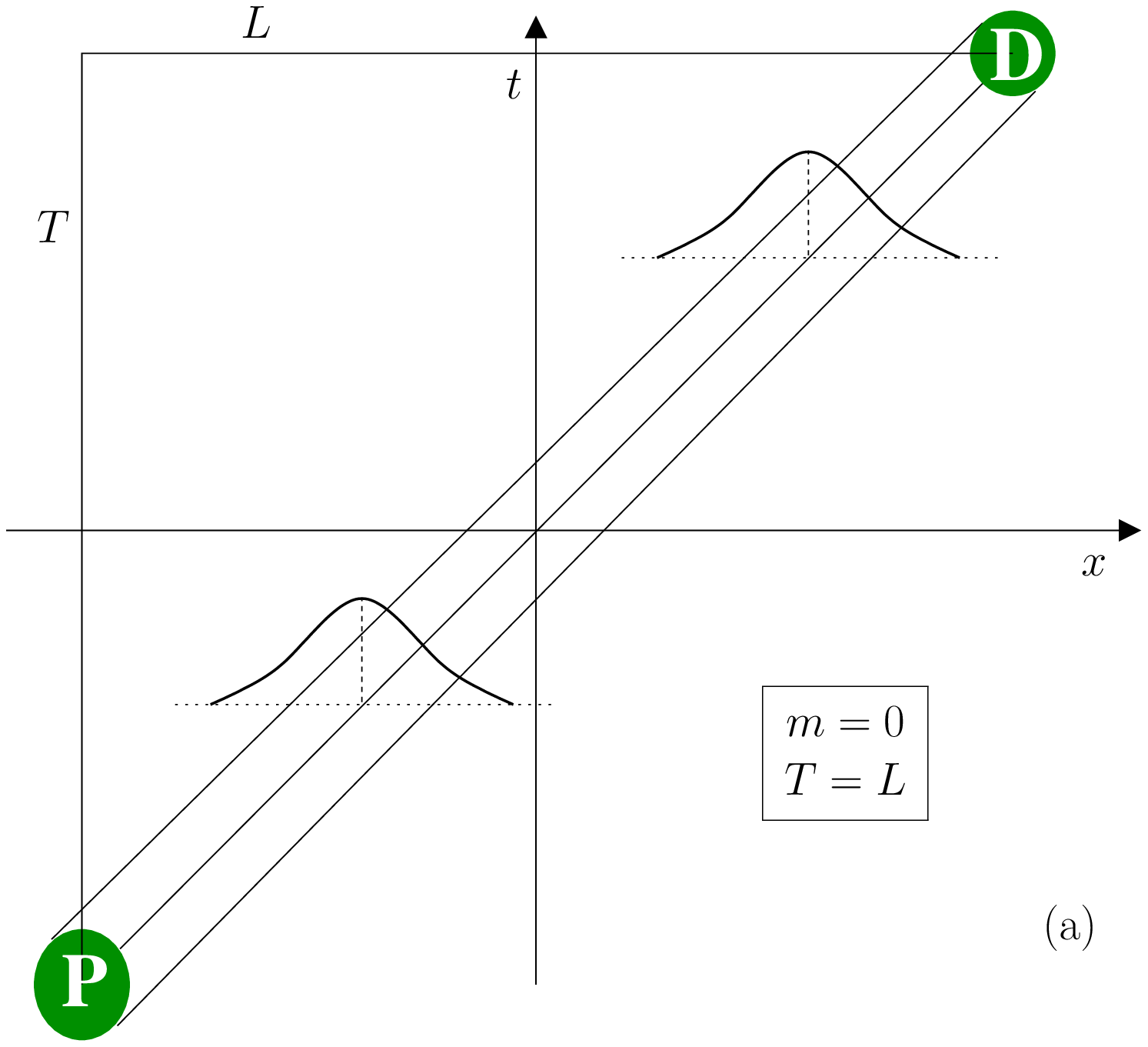}
\hspace{0.05\textwidth}
\includegraphics*[bb=84 361 511 770, width=0.45\textwidth]{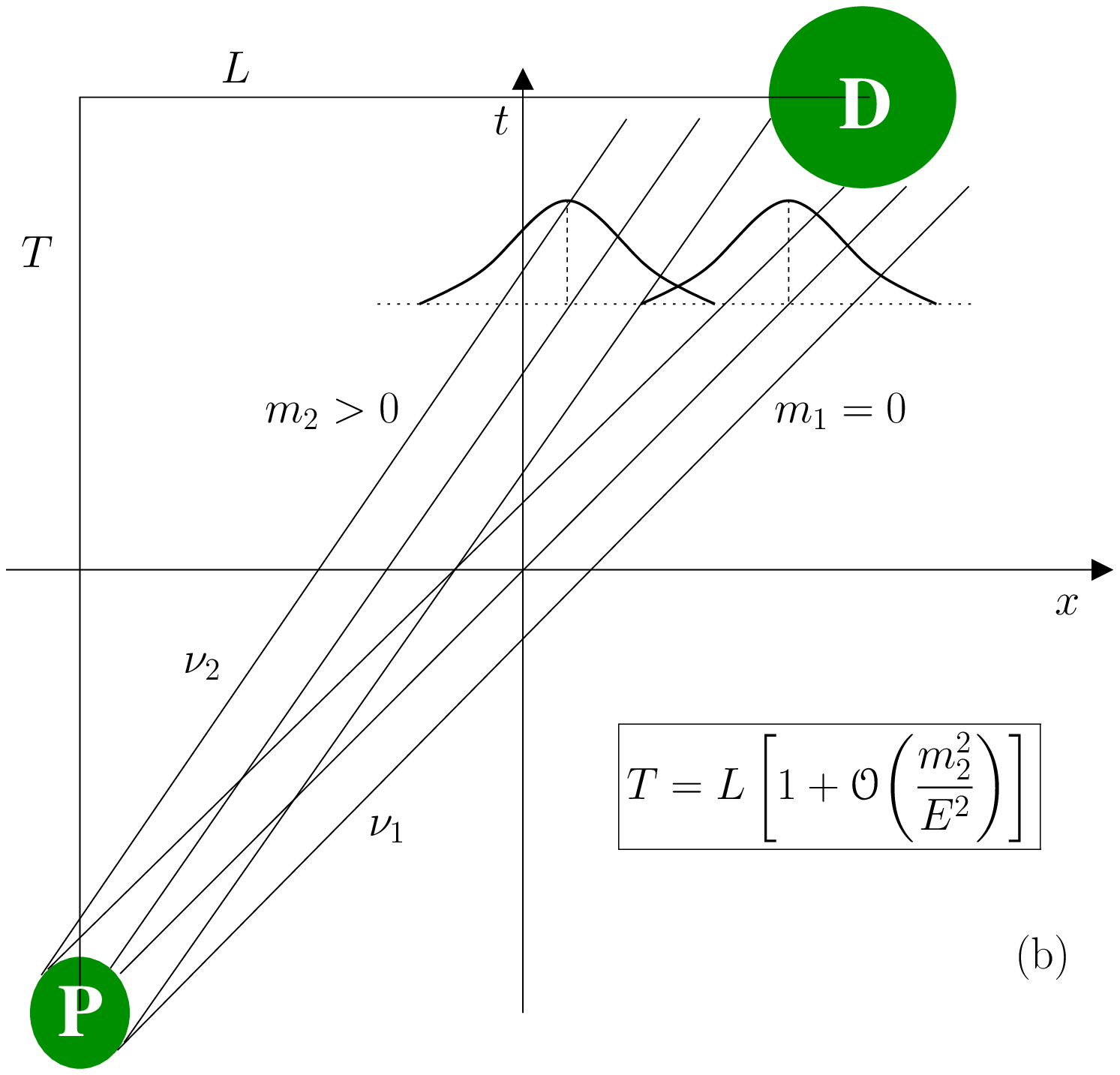}
\caption{ \label{h077}
Space-time diagram representing schematically
the propagation of the wave packet of a massless particle
(a) and the propagation of the wave packets of a superposition of
a massless and a massive ultrarelativistic particle (b)
from a production process P to a detection process D.
}
\end{figure}

The propagation of
a massless particle between localized production and detection processes
separated by $T \simeq L$
is illustrated schematically in the space-time diagram
in Fig.~\ref{h077}a.
The interesting case of propagation of a superposition of
two neutrinos with definite masses,
one massless ($\nu_{1}$) and one massive but ultrarelativistic ($\nu_{2}$)
is illustrated schematically in Fig.~\ref{h077}b.
One can note that in these diagrams both the production and detection processes
occupy a finite region in space-time,
called the \emph{coherence region},
in which the propagating particles are produced or detected coherently.
Indeed,
the uncertainty principle
implies that any interaction process $I$ has a space uncertainty
$\sigma_{x}^{\text{I}}$
related to the momentum uncertainty
$\sigma_{p}^{\text{I}}$
by
\begin{equation}
\sigma_{x}^{\text{I}} \, \sigma_{p}^{\text{I}} \sim \frac{1}{2}
\,.
\label{h079}
\end{equation}
A point-like process would have an infinite momentum uncertainty and
a process with definite momentum would be completely delocalized in space.
The momentum uncertainty can be estimated as the quadratic sum
of the uncertainties of the momenta of the localized particles taking part in the process:
\begin{equation}
(\sigma_{p}^{\text{I}})^{2}
\sim
\sum_{i} (\sigma_{p}^{i})^{2}
\,.
\label{h080}
\end{equation}
The sum is over the initial particles
and the final particles
which are localized through interaction with the environment.
Their momentum uncertainties $\sigma_{p}^{i}$ are related to the size
$\sigma_{x}^{i}$ of their wave packets by uncertainty relations
analogous to Eq.~(\ref{h079}),
\begin{equation}
\sigma_{x}^{i} \, \sigma_{p}^{i} \sim \frac{1}{2}
\,.
\label{h081}
\end{equation}
Therefore,
the space uncertainty of the process is given by
\begin{equation}
(\sigma_{x}^{\text{I}})^{-2}
\sim
\sum_{i} (\sigma_{x}^{i})^{-2}
\,.
\label{h082}
\end{equation}
It is clear that the particle with larger momentum uncertainty and associated
smaller space uncertainty gives the dominant contribution.

The coherence time $\sigma_{t}^{\text{I}}$ of an interaction process $I$
is the time over which the
wave packets of the interacting particles overlap.
If the process is the decay of a particle in vacuum,
the localization of such particle and its decay products is very poor and the
coherence time $\sigma_{t}^{\text{I}}$
is of the order of the particle lifetime.
On the other hand,
if the decay occurs in a medium where the decaying particle and its products
are well localized
or if the production process is a scattering process,
the coherence time can be estimated by
\begin{equation}
(\sigma_{t}^{\text{I}})^{-2}
\sim
\sum_{i}
\left( \frac{\sigma_{x}^{i}}{v_{i}} \right)^{-2}
\,,
\label{h083}
\end{equation}
where $v_{i}$ is the velocity of the particle $i$,
because $\sigma_{t}^{\text{I}}$
must be dominated by
the particle with smaller ratio
$\sigma_{x}^{i}/v_{i}$,
which is the first to leave the interaction region.
Therefore,
in general
$\sigma_{t}^{\text{I}} \gtrsim \sigma_{x}^{\text{I}}$,
in agreement with the physical expectation that
the coherence region of a process must be causally connected.

As illustrated in Fig.~\ref{h077},
one can estimate the size of the wave packet of a massive neutrino
created in a
production process P
as the coherence time $\sigma_{t}^{\text{P}}$
of the production process,
\begin{equation}
\sigma_{x}^{\nu}
\sim
\sigma_{t}^{\text{P}}
\,.
\label{h084}
\end{equation}
Let us emphasize that there is a profound difference between the
behavior of final neutrinos and other particles in the production process.
The initial particles
have wave packets which are determined by their creation process
or by previous interactions.
The initial particles and the final particles
which interact with the environment
contribute to the coherence time $\sigma_{t}^{\text{P}}$
through their
contribution to the momentum uncertainty in Eq.~(\ref{h080}).
An initial decaying particle contributes directly to
the coherence time $\sigma_{t}^{\text{P}}$
with its lifetime.
On the other hand, neutrinos are stable and leave the production process
without interacting with the environment.
Therefore,
they do not contribute to the determination
of the coherence time $\sigma_{t}^{\text{P}}$
and the size of their wave packets is determined by
$\sigma_{t}^{\text{P}}$.

Considering now the detection process $D$,
if there is only one particle propagating between the production and detection processes,
as shown in Fig.~\ref{h077}a,
the coherence size of the detection process is determined
by Eq.~(\ref{h080}),
with the sum over
all the participating particles which interact with the environment
and
the propagating particle,
which is described by a wave packet.
In the case of neutrino mixing,
the neutrino propagating between the production and detection processes
is in general a superposition of massive neutrino wave packets
which propagate with different phase velocity,
as illustrated in Fig.~\ref{h077}b.
In this case,
in the detection process,
the wave packets of different massive neutrinos
are separated by a distance
$ \Delta{x} = \Delta{v} \, T $,
where $\Delta{v}$ is the velocity difference.
If the source--detector distance is very large,
the separation of the massive neutrino wave packets at detection
may be larger than their size,
leading to the lack of overlap
\cite{Nussinov:1976uw}.
In this case,
the effective coherence size of the neutrino wave function
at the detection process is
\begin{equation}
\sigma_{x,\text{eff}}^{\nu}
\sim
\sqrt{
(\sigma_{x}^{\nu})^{2}
+
(\Delta{x})^{2}
}
\sim
\sqrt{
(\sigma_{t}^{\text{P}})^{2}
+
(\Delta{v} \, T)^{2}
}
\,.
\label{h085}
\end{equation}
However,
Eq.~(\ref{h082})
shows that
the particle with
smaller space uncertainty gives the dominant contribution to the
coherence size of the detection process.
Therefore,
if the
effective coherence size in Eq.~(\ref{h085}) of the neutrino wave function
is dominated by the separation of the wave packets
($\Delta{v} \, T \gg \sigma_{t}^{\text{P}}$)
and there is another particle participating in the detection
process which has much smaller space uncertainty,
the different massive neutrinos cannot be detected coherently.
In this case, there cannot be any interference between
the different massive neutrino contributions to the detection process
and the probability of transitions between different flavors
reduces to the incoherent transition probability
\begin{equation}
P_{\alpha\beta}^{\text{incoherent}}
=
\sum_{k}
|U_{{\alpha}k}|^{2}
\,
|U_{{\beta}k}|^{2}
\,,
\label{g034}
\end{equation}
which does not oscillate as a function of the source--detector distance.
On the other hand,
if all the other particles participating in the detection
process have space uncertainties which are larger than
effective coherence size in Eq.~(\ref{h085}) of the neutrino wave function,
the different massive neutrinos are detected coherently
\cite{hep-ph/9506271},
leading to the interference of their contributions
to the detection process
which manifests itself as oscillations of the probability
of flavor transitions, according to Eq.~(\ref{a001}).

These considerations show that
a wave-packet treatment of massive neutrinos
is important in order to understand
the coherence properties of neutrino oscillations.

\section{Quantum-Mechanical Wave-Packet Model}
\label{w01}

In this section we present a simple one-dimensional quantum-mechanical wave-packet model
\cite{Giunti:1991ca,hep-ph/9711363}
in which the momentum uncertainties of the states which describe the
produced and detected massive neutrinos
are approximated by Gaussian distributions.
More complete three-dimensional models in which
the neutrino momentum uncertainties are obtained from
a quantum field theoretical calculation of the
production and detection processes are discussed in
Refs.~\cite{hep-ph/9305276,hep-ph/9709494,hep-ph/9710289,hep-ph/9909332,hep-ph/0202068,hep-ph/0205014}.

Neglecting mass effects in the production and detection processes,
we describe the produced and detected neutrinos in a $\nu_{\alpha}\to\nu_{\beta}$ experiment
with the wave-packet flavor states
\begin{equation}
|\nu_{\alpha}^{\text{P}}\rangle
=
\sum_{k} U_{\alpha k}^* \int \mathrm{d}p \, \psi^{\text{P}}_{k}(p) \, |\nu_{k}(p)\rangle
\,,
\qquad
|\nu_{\beta}^{\text{D}}\rangle
=
\sum_{k} U_{\beta k}^* \int \mathrm{d}p \, \psi^{\text{D}}_{k}(p) \, |\nu_{k}(p)\rangle
\,,
\label{w02}
\end{equation}
with the Gaussian momentum distributions
\begin{equation}
\psi^{\text{P}}_{k}(p)
=
\frac{ 1 }{ \left( 2 \pi (\sigma_{p}^{\text{P}})^{2} \right)^{\frac{1}{4}} }
\exp\left[
- \frac{ \left( p - p_{k} \right)^{2} }{ 4 (\sigma_{p}^{\text{P}})^{2} }
\right]
\,,
\quad
\psi^{\text{D}}_{k}(p)
=
\frac{ 1 }{ \left( 2 \pi (\sigma_{p}^{\text{D}})^{2} \right)^{\frac{1}{4}} }
\exp\left[
- \frac{ \left( p - p_{k} \right)^{2} }{ 4 (\sigma_{p}^{\text{D}})^{2} }
\right]
\,.
\label{w03}
\end{equation}
The average momenta $p_{k}$ of the massive neutrinos are determined by the kinematics of
the production process.
They are the same in the detection process because of causality.
On the other hand, the energy-momentum uncertainties in the production and detection processes,
$\sigma_{p}^{\text{P}}$ and $\sigma_{p}^{\text{D}}$,
may be quite different.

The flavor transition amplitude is given by
\begin{align}
A_{\alpha\beta}(T,L)
=
\null & \null
\langle \nu_{\beta}^{\text{D}} |
\,
e^{ -i \mathsf{E} T + i \mathsf{P} L }
\,
| \nu_{\alpha}^{\text{P}} \rangle
\nonumber
\\
\propto
\null & \null
\sum_{k}
U_{\alpha k}^* \, U_{\beta k} \,
\int \mathrm{d}p \,
\exp\left[
- i E_{k}(p) T + i p L
- \frac{ \left( p - p_{k} \right)^{2} }{ 4 \sigma_{p}^{2} }
\right]
\,,
\label{w04}
\end{align}
with the massive neutrino energies
\begin{equation}
E_{k}(p)
=
\sqrt{ p^{2} + m_{k}^{2} }
\,.
\label{w041}
\end{equation}
and the global energy-momentum uncertainty
\begin{equation}
\frac{ 1 }{ \sigma_{p}^{2} }
=
\frac{ 1 }{ (\sigma_{p}^{\text{P}})^{2} }
+
\frac{ 1 }{ (\sigma_{p}^{\text{D}})^{2} }
\,.
\label{w05}
\end{equation}
This expression has a correct behavior from the physical point of view,
because the smaller energy-momentum uncertainty must dominate
in the determination of the total uncertainty.
On the other hand,
the global space-time uncertainty $\sigma_{x}=1/2\sigma_{p}$
is dominated by the largest of the space-time uncertainties
$\sigma_{x}^{\text{P}}=1/2\sigma_{p}^{\text{P}}$ and $\sigma_{x}^{\text{D}}=1/2\sigma_{p}^{\text{D}}$
of the production and detection processes:
\begin{equation}
\sigma_{x}
=
(\sigma_{x}^{\text{P}})^{2}
+
(\sigma_{x}^{\text{D}})^{2}
\,.
\label{h100}
\end{equation}

Since in practice the massive neutrino wave packets are always sharply peaked
at the average momentum ($
\sigma_{p}
\ll
E_{k}^{2}(p_{k}) / m_{k}
$),
we can approximate
\begin{equation}
E_{k}(p)
\simeq
E_{k} + v_{k} \left( p - p_{k} \right)
\,,
\label{w11}
\end{equation}
where $E_{k}$ and $v_{k}$ are, respectively,
the average energy and the group velocity given by
\begin{equation}
E_{k}
=
E_{k}(p_{k})
=
\sqrt{ p_{k}^{2} + m_{k}^{2} }
\,,
\qquad
v_{k}
=
\left.
\frac{ \partial E_{k}(p) }{ \partial p }
\right|_{p=p_{k}}
=
\frac{ p_{k} }{ E_{k} }
\,.
\label{w12}
\end{equation}
With this approximation,
the integration over $\mathrm{d}p$ in Eq.~(\ref{w04}) is Gaussian,
leading to
\begin{equation}
A_{\alpha\beta}(T,L)
\propto
\sum_{k}
U_{\alpha k}^* \, U_{\beta k} \,
\exp\left[
- i E_{k} T + i p_{k} L
- \frac{ \left( L - v_{k} T \right)^{2} }{ 4 \sigma_{x}^{2} }
\right]
\,.
\label{w13}
\end{equation}
Comparing with Eq.~(\ref{h030}), one can notice the additional
suppression factor for
$ | L - v_{k} T | \gtrsim \sigma_{x} $
due to the wave packets.

Finally, integrating the space-time dependent oscillation probability
$ P_{\alpha\beta}(T,L) = | A_{\alpha\beta}(T,L) |^{2} $
over the unobserved propagation time $T$,
we obtain, for ultrarelativistic neutrinos,
\begin{equation}
P_{\alpha\beta}(L)
=
\sum_{k,j}
U_{\alpha k}^* \, U_{\beta k} \,
U_{\alpha j} \, U_{\beta j}^* \,
\exp\left[
- 2 \pi i \, \frac{ L }{ L^{\mathrm{osc}}_{kj} }
-
\left( \frac{ L }{ L^{\mathrm{coh}}_{kj} } \right)^{2}
-
2 \pi^{2} \xi^{2} \left( \frac{ \sigma_x }{ L^{\mathrm{osc}}_{kj} } \right)^{2}
\right]
\,,
\label{w14}
\end{equation}
with the oscillation and coherence lengths
\begin{equation}
L^{\mathrm{osc}}_{kj}
=
\frac{4 \pi E}{\Delta{m}^{2}_{kj}}
\,,
\qquad
L^{\mathrm{coh}}_{kj}
=
\frac{4 \sqrt{2} E^{2}}{|\Delta{m}^{2}_{kj}|}
\,
\sigma_{x}
\,.
\label{w15}
\end{equation}
The coefficient $\xi$,
which is the only quantity in Eq.~(\ref{w14}) depending on the production process,
comes from the general ultrarelativistic approximation
\cite{Giunti:1991ca,Giunti:2000kw,Giunti:2001kj,Giunti:2002xg,hep-ph/0608070,Giunti-Kim-2007}
\begin{equation}
p_{k}
\simeq
E
-
\left( 1 - \xi \right) \frac{m_{k}^{2}}{2E}
\,,
\qquad
E_{k}
\simeq
E
+ \xi \,\frac{m_{k}^{2}}{2E}
\,.
\label{w16}
\end{equation}

In the limit of
negligible wave packet effects,
i.e.\ for
$L \ll L^{\text{coh}}_{kj}$
and
$\sigma_{x} \ll L^{\text{osc}}_{kj}$,
the oscillation probability in the wave packet approach reduces to the standard one
in Eq.~(\ref{a001}),
obtained in the plane wave approximation.
The additional
\emph{localization} and \emph{coherence} terms
\begin{equation}
P_{kj}^{\text{loc}}
=
\exp\left[
-
2 \pi^{2} \xi^{2} \left( \frac{ \sigma_x }{ L^{\mathrm{osc}}_{kj} } \right)^{2}
\right]
\,,
\qquad
P_{kj}^{\text{coh}}
=
\exp\left[
-
\left( \frac{ L }{ L^{\mathrm{coh}}_{kj} } \right)^{2}
\right]
\,,
\label{h125}
\end{equation}
have the following physical meaning
\cite{Giunti:1991ca,hep-ph/9711363,hep-ph/0109119,hep-ph/0205014,hep-ph/0302026,hep-ph/9305276,hep-ph/9709494,hep-ph/9909332,hep-ph/0202068,hep-ph/0402217,hep-ph/0608070}.

The localization term $P_{kj}^{\text{loc}}$
suppresses the oscillations due to
$\Delta{m}^{2}_{kj}$
if
$ \sigma_{x} \gtrsim L^{\text{osc}}_{kj} $.
This means that in order to measure the interference
of the massive neutrino components
$\nu_{k}$ and $\nu_{j}$
the production and detection processes must be localized
in space-time regions much smaller than the
oscillation length $L^{\text{osc}}_{kj}$.
In practice this requirement is satisfied in all
neutrino oscillation experiments.

The localization term allows one to distinguish
neutrino oscillation experiments
from experiments on the measurement of neutrino masses.
As first shown in Ref.~\cite{Kayser:1981ye},
neutrino oscillations are suppressed
in experiments which are
able to measure,
through energy--momentum conservation,
the mass of the neutrino.
Indeed,
from the energy--momentum dispersion relation in Eq.~(\ref{h029})
the uncertainty of the mass determination is
\begin{equation}
\delta{m_{k}}^{2}
=
\sqrt{
\left( 2 \, E_{k} \, \delta{E_{k}} \right)^{2}
+
\left( 2 \, p_{k} \, \delta{p_{k}} \right)^{2}
}
\simeq
2 \sqrt{2} \, E \, \sigma_{p}
\,,
\label{h127}
\end{equation}
where the approximation holds for ultrarelativistic neutrinos.
If
$
\delta{m_{k}}^{2}
<
|\Delta{m}^{2}_{kj}|
$,
the mass of $\nu_{k}$ is measured
with an accuracy better than the difference
$\Delta{m}^{2}_{kj}$.
In this case
the neutrino $\nu_{j}$ is not produced or detected
and the interference of
$\nu_{k}$ and $\nu_{j}$
which would generate oscillations does not occur.
The localization term
automatically
suppresses
the interference of
$\nu_{k}$ and $\nu_{j}$,
because
\begin{equation}
-
2 \pi^{2}
\left(
\frac{ \sigma_{x} }{ L^{\text{osc}}_{kj} }
\right)^{2}
=
-
\left(
\frac{ \Delta{m}^{2}_{kj} }{ 4 \sqrt{2} E \sigma_{p} }
\right)^{2}
\simeq
- \frac{1}{4}
\left(
\frac{ \Delta{m}^{2}_{kj} }{ \delta{m_{k}}^{2} }
\right)^{2}
\,.
\label{h128}
\end{equation}

If the condition
\begin{equation}
\sigma_{x} \ll L^{\text{osc}}_{kj}
\,,
\label{h129}
\end{equation}
which is
necessary for unsuppressed interference of $\nu_{k}$ and $\nu_{j}$,
is satisfied,
as usual in neutrino oscillation experiments,
the localization term can be neglected,
leading to the flavor transition probability
\begin{equation}
P_{\alpha\beta}(L)
=
\sum_{k,j}
U_{\alpha k}^{*} U_{\alpha j} U_{\beta k} U_{\beta j}^{*}
\exp\left[
- 2 \pi i \frac{ L }{ L^{\text{osc}}_{kj} }
-
\left(
\frac{ L }{ L^{\text{coh}}_{kj} }
\right)^{2}
\right]
\,,
\label{h130}
\end{equation}
which is a function of the distance $L$,
depending on the oscillation and coherence lengths in Eq.~(\ref{w15}).

\begin{figure}[t!]
\includegraphics*[bb=72 679 546 757, width=\textwidth]{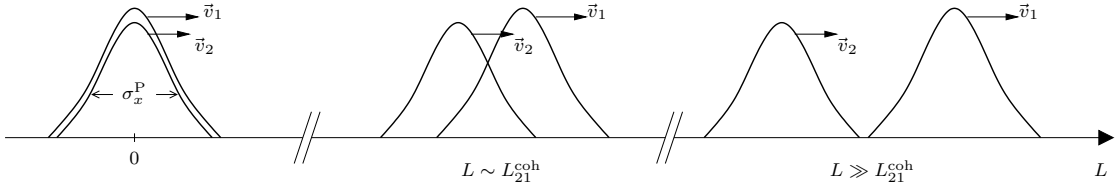}
\caption{ \label{h124}
Schematic illustration of the separation
of two wave packets with different group velocities,
produced coherently
at $L=0$ with widths $\sigma_{x}^{\text{P}}$
determined by the coherence size of the production process.
The coherence size of the detection process is assumed to be negligible.
}
\end{figure}

In Eq.~(\ref{h130}),
each $k,j$ term contains,
in addition to the standard oscillation phase,
the coherence term $P_{kj}^{\text{coh}}$,
which suppresses the interference of the massive neutrinos
$\nu_{k}$ and $\nu_{j}$
for distances
larger than the corresponding coherence length,
i.e.\ for
$ L \gg L^{\text{coh}}_{kj}$.
This suppression is due to the separation of the
different massive neutrino wave packets,
which propagate with different velocities,
as illustrated in Figs.~\ref{h077}b and \ref{h124}.
When the wave packets of
$\nu_{k}$ and $\nu_{j}$
are so much separated that they cannot both overlap with the detection process,
the massive neutrinos
$\nu_{k}$ and $\nu_{j}$
cannot be absorbed coherently
\cite{Nussinov:1976uw,hep-ph/9506271}.
In this case, only one of the two massive neutrinos contributes
to the detection process and
the interference effect which produces the oscillations is absent.
However,
in general,
the flavor transition probability does not vanish.
For example,
if
$ L \gg L^{\text{coh}}_{kj}$
for all $k$ and $j$,
the flavor transition probability reduces to the incoherent transition probability in Eq.~(\ref{g034}).

\section{Conclusions}
\label{Conclusions}

We have reviewed the standard theory of neutrino oscillations,
highlighting the three main assumptions:
\ref{A1}
the definition of the flavor states,
\ref{A2}
the equal-momentum assumption
and
\ref{A3}
the time $=$ distance assumption.

We have shown that the
flavor neutrino state that describes a neutrino produced or detected in
a charged-current weak interaction process
depends on the process under consideration.
The standard flavor states are
correct approximations of these states in oscillation experiments,
which are not sensitive to the
dependence of neutrino interactions on the different neutrino masses.

We have presented a covariant plane-wave theory of neutrino oscillations
in which both the evolutions in space and in time
of the neutrino state are taken into account,
leading to the standard probability of flavor transitions.
In this model,
no assumption on the energies and momenta of the propagating massive neutrinos is needed.
Moreover,
the derivation of the Lorentz-invariant flavor transition probability
is manifestly Lorentz invariant.

We have argued that the time $=$ distance assumption derives from the wave-packet character
of the propagating neutrinos.
We have discussed the necessity of a wave-packet treatment of neutrino oscillations
for the description of the localization of the production and detection processes
and the coherence of the oscillations.
We have also presented a simple quantum-mechanical wave-packet model
which leads to the standard probability of flavor transitions
with additional localization and coherence terms which have
important physical meaning.

In conclusion,
we would like to emphasize that the insight of the founders of
the theory of neutrino oscillations led them to the correct standard expression for the
flavor transition probability.
Our more modest task has been to clarify the assumptions
and to try to improve the derivation
hoping to elucidate the deep physical nature of neutrino oscillations.


\begin{theacknowledgments}
I would like to thank the Department of Theoretical Physics of the University of Torino
for hospitality and support.
\end{theacknowledgments}

\end{document}